\documentclass[graybox]{svmult}

\usepackage{mathptmx}       
\DeclareMathAlphabet{\mathcal}{OMS}{cmsy}{m}{n}
\usepackage{helvet}         
\usepackage{courier}        
\usepackage{type1cm}        
\usepackage{makeidx}         
\usepackage{graphicx}        
\usepackage{verbatim}
\usepackage{multicol}        
\usepackage[bottom]{footmisc}

\usepackage{multirow}
\usepackage{natbib}
\usepackage{graphicx}
\usepackage{makecell}
\usepackage{booktabs}
\usepackage{array}
\usepackage{url}
\usepackage{algorithm}
\usepackage{algorithmic}
\usepackage{dsfont}
\usepackage{mathtools}
\usepackage{enumitem}
\usepackage{mdframed}
\usepackage{color,xcolor}
\usepackage[OT1,T3,T1]{fontenc}
\usepackage{amsfonts}

\def\1{\bm{1}}

\newcommand{\ind}{\mathds{1}}

\newcommand{\Xb}{\mathbf{X}}

\newcommand{\cA}{\mathcal{A}}

\newcommand{\cH}{\mathcal{H}}
\newcommand{\cI}{\mathcal{I}}

\newcommand{\cM}{\mathcal{M}}
\newcommand{\cN}{\mathcal{N}}

\newcommand{\cR}{\mathcal{R}}

\newcommand{\cT}{{\mathcal{T}}}

\newcommand{\EE}{\mathbb{E}}

\newcommand{\PP}{\mathbb{P}}

\makeindex             

\definecolor{NavyBlue}{RGB}{51,112,185}

\begin{document}
\title*{False Discovery Control in Multiple Testing:\\ A Brief Overview of Theories and Methodologies}
\titlerunning{A Brief Overview of Multiple Testing Procedures}
\author{Jianliang He, Bowen Gang, and Luella Fu}
\institute{Jianliang He\at Department of Statistics and Data Science, Fudan Univeristy, \email{hejl20@fudan.edu.cn}.
\and Bowen Gang \at Department of Statistics and Data Science, Fudan Univeristy. \email{bgang@fudan.edu.cn}.
\and Luella Fu \at Department of Mathematics, San Francisco State University. \email{luella@sfsu.edu}.}

%
%
\maketitle
\abstract{	As the volume and complexity of data continue to expand across various scientific disciplines, the need for robust methods to account for the multiplicity of comparisons has grown widespread. A popular measure of type 1 error rate in multiple testing literature is the false discovery rate (FDR). The FDR provides a powerful and practical approach to large-scale multiple testing and has been successfully used in a wide range of applications. The concept of FDR has gained wide acceptance in the statistical community and various methods has been proposed to control the FDR. In this work, we review the latest developments in FDR control methodologies. We also develop a conceptual framework to better describe this vast literature; understand its intuition and key ideas; and provide guidance for the researcher interested in both the application and development of the methodology.}

\section{Introduction}
In modern scientific research, massive and complex data sets with thousands or even millions of variables are collected routinely by governments, researchers, organizations, small businesses, and large enterprises.
If not adequately addressed, multiplicity in large-scale studies may precipitate substantial issues in the reproducibility of findings, the emergence of publication bias, and the susceptibility to p-hacking within the scientific research community \citep{ioannidis2005most,head2015extent}. \cite{goodman2016does} claimed that multiplicity, combined with incomplete reporting, might be the largest contributor to the non-reproducibility or falsity of published claims. 

Potential issues in error rate control arising from multiplicity have long been acknowledged. In 1953, to address the problem of multiplicity,  John Tukey introduced the notion of family-wise error rate (FWER) which is the probability of making one or more false discoveries.  When the number of hypotheses under consideration is large, FWER is usually too restrictive. To increase testing power, it is more reasonable to control the false discovery rate (FDR) \citep{benjamini1995controlling}, which is defined as the expected proportion of false discoveries among all discoveries. Since its introduction in 1995, FDR has become a central concept in modern statistics, and
the literature has grown beyond what can reasonably be synthesized in this review. The choice of topics and cited references in our work is not intended to serve as an appraisal of their respective importance or contributions. Instead of striving for an exhaustive survey, which is an impractical endeavor, we have deliberately focused on a selection of subjects that we believe can be articulated with clarity and are likely to be highly relevant to those in the field of practice.

In this review, we give a generic framework for FDR control and show how some of the most popular FDR methods conform to this framework. We also consider methods that incorporate auxiliary information and methods that are valid under dependence. 
Both frequentist and Bayesian perspectives are considered.

In Section 2, we formally present the  FDR control problem for multiple testing. In Section 3, we discuss the celebrated Benjamini-Hochberg (BH) procedure \citep{benjamini1995controlling} and its variations. In Section 4, we review the Sun-Cai (SC) procedure \citep{sun2007oracle} and its variations. In Section 5, the discussion is centered on the control of the FDR in the context of dependent data. Concluding the review in Section 6, we identify and highlight further topics pertinent to FDR control that may be of interest to the reader.
\textbf{Notation.} Throughout the paper, we denote $\cM$ or $[m]=\{1,\dots,m\}$  as the index set of all hypotheses being simultaneously tested. Let $\ind(\cdot)$ denote the indicator function that returns 1 if the condition in $(\cdot)$ is satisfied and 0 otherwise, and $|\cA|$ denotes the cardinality of set $\mathcal{A}$. We use $a\vee b$ to denote =$\max\{a,b\}$.  
\section{Problem Formulation}\label{sec:formulation}
	Suppose we want to test $m$ null hypotheses $H_{0,1},\dots,H_{0,m}$ simultaneously and are given corresponding summary statistics $X_1,\dots,X_m$. Further assume the null distributions of the summary statistics are known. For example, if $X_1,\dots,X_m$ are p-values we assume they follow ${\rm Unif}[0,1]$ under the null. 
	A multiple testing procedure examines these summary statistics and decides which null hypotheses to reject. Let $\cH_0=\{i:H_{0,i}\ \text{is true}\}$ be the index set of true nulls hypotheses, $\cH_1=\cM/\cH_0$ as the index set of non-null hypotheses. Denote $\delta_i$ the decision for $H_{0,i}$ with $\delta_i=0$ indicating non-rejection of  $H_{0,i}$ and $1$ indicating rejection. Denote $\cR=\{i:H_{0,i}\ \text{is rejected}\}$ the rejection set.  Denote $\theta_i=\ind(i\in\cH_0)$ the true state of the $i$-th hypotheses. 
	A selection error, or false positive, occurs if the practitioner asserts that $H_{0,i}$, is false when it is not. In multiple testing problems, such false positive decisions are inevitable if we wish to discover interesting effects with reasonable power. Instead of aiming to avoid any false positives, a practical goal is to keep the false discovery rate (FDR) small. The FDR is the expectation of the false discovery proportion (FDP) among all rejections:
	\begin{equation}
		\label{eq:fdr}
		\text{FDR}(\cR) = \mathbb E\left[\text{FDP}(\cR)\right]~\text{where FDP}(\cR) = \dfrac{|\cR\cap\cH_0|}{|\cR|\vee 1}.
	\end{equation}

We use the true discovery rate (TDR) to measure the power of a testing procedure. It is defined as
the expectation of true discovery proportion.
	\begin{equation}
		\label{eq:tdr}
		\text{TDR}(\cR) = \mathbb E\left[\text{TDP}(\cR)\right]~\text{where TDP}(\cR) = \dfrac{|\cR\cap\cH_1|}{|\cH_1|\vee 1}.
	\end{equation}
	The goal is to devise a testing procedure that has high TDR while ensuring that FDR is below a pre-determined level $\alpha$.
	
		We mention two popular variants of FDR. One is 
		 is the marginal FDR (mFDR) \citep{storey2002direct} defined as
	\begin{equation}
		\label{eq:mfdr}
		\text{mFDR}(\cR) = \dfrac{\mathbb E\left[|\cR\cap\cH_0|\right]}{\mathbb E\left[|\cR|\right]}.
	\end{equation}
	It was shown in \cite{genovese2002operating} that under weak conditions mFDR and FDR are asymptotically equivalent. We provide the precise statement in the next proposition.

	\begin{proposition}
		Let $\cR$ be the rejection set concerning $m$ null hypotheses following a specific decision rule, then ${\rm FDR}(\cR)={\rm mFDR}(\cR)+o(1)$ as $m\rightarrow\infty$ if the following two conditions hold: 1) there exists an absolute constant $\eta>0$ such that $m^{-1}\mathbb{E}[|\cR|]\geq\eta$, 2) ${\rm Var}(|\cR|)=o(m^2)$.
		\label{prop:fdrmfdr}
	\end{proposition}
One attractive feature of mFDR is its ability to be "combined." Specifically, if we apply two testing rules that control mFDR at level $\alpha$ on two separate sets of null hypotheses, then the overall mFDR on the union of the two sets of null hypotheses is still controlled at level $\alpha$. However, \cite{benjamini1995controlling} points out that it is sometimes impossible to control mFDR. One such setting is where all null hypotheses are true, in which case the mFDR will either be undefined or equal to 1.

Another popular variant of FDR is the positive false discovery rate (pFDR), defined as:

 $$
 \text{pFDR}(\cR) = \mathbb E\left[\text{FDP}(\cR)|\cR>0\right]=\text{FDR}(\cR)/\mathbb{P}(\cR>0) .
 $$
pFDR has a natural Bayesian interpretation, as illustrated in \cite{storey2003positive}. It is also shown in \cite{storey2003positive} that under some weak assumptions, pFDR is equivalent to mFDR.

	\subsection{A Generic Framework for FDR Control}\label{sec:framework} 
We first outline a generic framework for FDR control. The vast majority of methodologies developed to control FDR in offline analyses adhere to the three--step framework outlined below.
	
	\begin{enumerate}[leftmargin=30pt]
		\item[\textbf{Step 1}] \textbf{(Ranking)} Construct a suitable summary statistics $T_i$ for for each $H_{0,i}$ and rank the null hypotheses according to $T_i$.
		\item[\textbf{Step 2}] \textbf{(FDP Estimation)} For any given $t$ estimate the FDP of the decision rule $\pmb{\delta}(t)=\{\delta_1(t),\ldots, \delta_m(t)\}$, where $\delta_i(t)=\ind(T_i\leq t)$. Denote the estimate as $\widehat{\rm FDP}(t)$.
		\item[\textbf{Step 3}] \textbf{(Thresholding)} For a given target FDR level $\alpha$, define
		$
		t_\alpha=\sup\{t\in\cT:\widehat{\rm FDP}(t)\leq\alpha\},
		$. Reject $H_{0,i}$ if and only if $T_i\leq t_\alpha$.
		\end{enumerate}
		Innovation within the FDR literature predominantly resides in the formulation of more informative summary statistics (Step 1) and the design of sharper estimators for the FDP (Step 2). 
		In the next two sections, we examine several prominent methodologies for FDR control, elucidating their alignment with the three-step framework delineated above.
		

\section{ The BH procedure and its variations}\label{sec:convention}
 Introduced by \cite{benjamini1995controlling}, the BH procedure inaugurated the control of the   false discovery rate (FDR) and continues to be one of the most seminal and prevalently applied methodologies in multiple hypothesis testing. In this section we provide an overview of the BH procedure and its variations.

\subsection{The BH procedure }\label{sec:BH}
 The BH procedure takes p-values $P_1,\ldots,P_m$ and target FDR level $\alpha$ as inputs. It computes threshold $P_{(k)}$ where
$$
k=\max\left\{j: \ P_{(j)}\leq \frac{\alpha j}{m}\right\},\quad P_{(1)}\leq \ldots\leq P_{(m)}.
$$
 Then it rejects $H_{0,i}$ if and only if $P_i\leq P_{(k)}$.
 In the framework of Section \ref{sec:framework}, 
 the BH procedure uses the p-value as the ranking statistic and estimates the FDP at threshold $t$ by
 	\begin{equation}\label{fdp:bh}
 	\widehat{\rm FDP}(t)=\frac{mt}{\sum_{i=1}^m\ind(P_i\leq t)}.
 \end{equation}
 The denominator in \eqref{fdp:bh} is the number of hypotheses rejected at threshold $t$. Note that if $P_{(j)}$ is the threshold then $\sum_{i=1}^m\ind(P_i\leq P_{(j)})=j$. The numerator is the estimated number of false rejections in the worst case where all the null hypotheses are true. The theoretical guarantee of the BH procedure is given in the next theorem.
 \begin{theorem}\label{thm:BH}
 Assume the p-values are independent under the nulls. Then  BH procedure controls FDR at level $\alpha |\cH_0|/m$.
 \end{theorem}
 There are two elementary proofs for Theorem \ref{thm:BH} (details are in \cite{wang2022elementary}):  
 one proof employs the concept of martingales, while the other utilizes a technique known as ``leave-one-out."  
 Both proofs are fundamental, serving as pivotal inspirations for the development of numerous sophisticated testing procedures. 


\subsection{$\pi_0$-Adaptive BH Procedure}\label{sec:adaBH}
 From Theorem \ref{thm:BH} we see that the BH procedure tends to be conservative as it controls FDR at level $(|\cH_0|/m)\cdot\alpha$ instead of $\alpha$.
 One remedy for this conservativeness is to estimate the null proportion $\pi_0:=|\cH_0|/m$ and replace 
the target FDR level in the input of BH procedure by $\alpha/\hat{\pi}_0$. The problem of estimating $\pi_0$ has been discussed extensively in the literature \citep[e.g.,][]{storey2002direct, meinshausen2006estimating, jin2007estimating,chen2019uniformly}. The most popular estimator of $\pi$ is the one proposed in \cite{storey2004strong}, which is computed as 
\begin{equation}
	\hat{\pi}_0(\lambda)=\frac{1+\sum_{i=1}^m\ind(P_i\geq\lambda)}{m(1-\lambda)}.
	\label{eq:storeypi}
\end{equation}
 The intuition behind Storey's null proportion estimator $\hat{\pi}_0(\lambda)$ in \eqref{eq:storeypi} is that  
\begin{align}\label{eq1}
	\hat{\pi}_0(\lambda)=\frac{1+\sum_{i=1}^m\ind(P_i\geq\lambda)}{m(1-\lambda)}&\geq \frac{\sum_{i\in\cH_0}\ind(P_i\geq\lambda)}{|\cH_0|\cdot(1-\lambda)}\cdot\frac{|\cH_0|}{m}\\
	&\approx\pi_0\cdot\frac{\PP(P_i\geq\lambda\mid i\in\cH_0)}{1-\lambda}=\pi_0,\nonumber
\end{align}
where "$\approx$" comes from the law of large numbers. We can see from the above calculation that 
 $\hat{\pi}_0(\lambda)$ tends to overestimate $\pi_0$ for any choice of $\lambda$. This is desirable as it makes the testing procedure conservative. 
 There is an inherent bias-variance trade-off in the choice of $\lambda$. The ``$\geq$" in \eqref{eq1} is sharper when $\lambda$ is bigger. However, a bigger $\lambda$ also results in larger variance for $\hat{\pi}_0(\lambda)$. 
As $\lambda$ shrinks, the bias of $\hat{\pi}_0(\lambda)$ grows larger, but the variance becomes smaller. The choice of proper $\lambda$ has been studied in the literature. For example \cite{storey2002direct,storey2003positive} proposed a bootstrapping and a spline-smoothing method, respectively. \cite{langaas2005estimating} developed a class of estimators based on nonparametric maximum likelihood estimates (MLEs). 

For any given $\lambda$, we can estimate the FDP by \begin{equation}\label{fdpbh}
 \widehat{\rm FDP}_\lambda(t)=\frac{\hat{\pi}_0(\lambda)\cdot mt}{\sum_{i=1}^m\ind(P_i\leq t)}.
\end{equation}
Define $t^\lambda_\alpha=\sup\{t\in(0,1-\lambda]:\widehat{\rm FDP}_\lambda(t)\leq\alpha\}$. It was shown in \cite{storey2004strong} that the procedure which rejects $H_{0,i}$ if and only if $p_i\leq t^\lambda_\alpha$ asymptotically controls FDR at the target level. 
Recently, \cite{gao2023adaptive} proposed a new framework for choosing $\lambda$ in a data-dependent fashion and demonstrated that the associated BH procedure controls the FDR in finite samples.
\subsection{Mirror Sequence and Statistic}
The problem of determining $\lambda$ can be avoided entirely if we allow $\lambda$ to vary. To see this recall that
$$
 \widehat{\rm FDP}_\lambda(t)=\frac{\hat{\pi}_0(\lambda)\cdot mt}{\sum_{i=1}^m\ind(P_i\leq t)}=\frac{1+\sum_{i=1}^m\ind(P_i\geq\lambda)}{m(1-\lambda)}\dfrac{mt}{\sum_{i=1}^m\ind(P_i\leq t)}.
$$
Suppose  we take $\lambda=1-t$ then the above reduces to
\begin{equation}\label{fdpbc}
	\widehat{\rm FDP}_{1-t}(t)=\dfrac{1+\sum_{i=1}^{m}\ind(1-P_i\leq t)}{\sum_{i=1}^{m}\ind(P_i\leq t)}.
\end{equation}
In essence, \eqref{fdpbc} is exploiting the symmetry of $1-P_i$ and $P_i$ as both are ${\rm Unif}[0,1]$ under the null.  For this reason $1-P_i$ is sometimes called the \textit{mirror sequence} \citep{leung2022zap}. 
The methods developed in \cite{lei2018adapt,leung2022zap}, and \cite{zhang2022covariate} all rely on this crucial idea. If \eqref{fdpbc} is used to estimate the FDP, the threshold should be modified as
\begin{equation}\label{BC}
t_\alpha=\sup\{t\in(0,0.5]:\widehat{\rm FDP}_{1-t}(t)\leq\alpha\}.
\end{equation}We have the following theoretical guarantee, the proof of which uses martingale theory. Details can be found in \cite{leung2022zap}.
 \begin{theorem}\label{thm:BC}
	The procedure that rejects $H_{0,i}$ if and only if $P_i\leq t_\alpha$ where $t_\alpha$ is as defined in \eqref{BC} controls FDR at level $\alpha$.
\end{theorem}

The technique of using symmetry to estimate FDP is not restricted to mirror sequence.
Another approach is to utilize \text{mirror statistic}. While mirror sequences rely on the pairing symmetry of existing test statistics such as $P_i$ and $1-P_i$, mirror statistics involve the transformation or creation of test statistics to exhibit mirror-like properties.
\cite{barber2015controlling, barber2019knockoff} introduced the ``knockoff" technique for FDP estimation within the framework of variable selection in regression analysis. The key idea is to construct ``knockoff“ variants of the initial variables that maintain the intrinsic correlation structure of the original dataset while ensuring the knockoffs have no association with the response variable. Consequently, the distribution of the coefficients corresponding to these knockoff variables is symmetric about zero. Employing this symmetry, the knockoff filter is able to control the FDR. The idea of constructing mirror statistics and use that symmetry to control FDR was also explored in \cite{dai2023false} and \cite{du2023false}.

\subsection{Weighted BH procedure}
In many applications relevant domain knowledge, such as carefully constructed auxiliary sequences from the same dataset \citep{liu2014incorporation,tony2019covariate} and external covariates or prior data from secondary data sources \citep{fortney2015genome,scott2015false,ignatiadis2016data,zhang2022covariate}, are often available alongside the primary dataset. These elements of prior knowledge can be exploited to improve the performance of testing procedures.
 A pragmatic approach to integrating such prior knowledge is  p-value weighting. This technique adjusts the significance of each test based on a priori information, thereby enhancing the sensitivity and interpretability of the overall testing procedure. p-value weighting was studied by \cite{genovese2006false,cai2022laws,durand2019adaptive}; and many others. 

The generic framework for p-value weighting is as follows: given positive weights $w_1,\ldots, w_m$ that are independent of $P_1,\ldots,P_n$ conditional on $\theta_1,\ldots,\theta_m$, compute the weighted p-values $Q_1,\ldots,Q_m$ where $Q_i=P_i/w_i$. We consider procedures that reject $H_{0,i}$ if and only if $Q_i\leq t$ for some $t$. For such procedures, the FDP can be estimated by
\begin{equation}\label{eq3}
	\widehat{\rm FDP}^{\rm wBH}(t)=\dfrac{\sum_{i=1}^{m}w_it}{\sum_{i=1}^{m}\ind(Q_i\leq t)}.
\end{equation}
$H_{0,i}$ is rejected if and only if $Q_i\leq t_\alpha$ where $t_\alpha=\max\{t:	\widehat{\rm FDP}^{\rm wBH}(t)\leq \alpha\}$. When the weighted BH procedure was first studied in \cite{genovese2006false} the authors set   $\sum_{i=1}^{m}w_i\leq m$. Under this condition, the aforementioned procedure  is equivalent to applying the BH procedure to the weighted p-values $Q_1,\ldots, Q_m$, hence the name. It was shown in \cite{genovese2006false} that the weighed BH procedure controls the FDR at level $\alpha$.

In numerous instances, the weights are not provided explicitly; instead, side information denoted as $s_i$ accompanies each p-value $P_i$. This auxiliary data can include various types of domain-specific knowledge, such as gene expression levels in genomic studies, demographic information in clinical trials, or historical performance metrics in economic forecasting. The motivation for using this side information to define weights lies in its potential to improve the power and accuracy of the BH procedure by incorporating relevant contextual information. For example, in genomic studies, genes known to be involved in certain biological pathways might be assigned higher weights, reflecting their higher prior probability of being significant.

To integrate this side information into the testing procedure, \cite{cai2022laws,li2019multiple} suggested constructing weights that reflect the prior probabilities
 ${\pi}_0(s_i):=\mathbb{P}(\theta_i=0|s_i)$. One way of estimating ${\pi}_0(s_i)$ is to use a generalization of \eqref{eq:storeypi}
\begin{equation}\label{pihat}
\hat{\pi}_0(s)=\dfrac{\sum_{i=1}^{m}K_h(s_i-s)\ind(P_i>\lambda)}{(1-\lambda)\sum_{i=1}^{m}K_h(s_i-s)},
\end{equation}
where $K_h(t)=h^{-1}K(t/h)$ and $K$ a kernel function satisfies
$\int K(t)dt=1$, $\int tK(t)dt=0$ and $\int t^2K(t)dt<\infty$. \cite{li2019multiple} proposed using  $1/(1-\hat{\pi}_0(s_i))$ as weights, whereas \cite{cai2022laws} recommended using $\hat{\pi}_0(s_i)/(1-\hat{\pi}_0(s_i))$ instead to mimic the optimal ranking. \cite{ignatiadis2024values} proposed using e-values (see Definition \ref{eval}) as weights. It is shown in \cite{ignatiadis2024values} that if the $w_i$'s are e-values, then the condition $\sum_{i=1}^{m} w_i \leq m$ is no longer necessary for the validity of the weighted BH procedure.

\section{The SC procedure and its variations}\label{sec:Bayesian}
The BH procedure and its variations take a frequentist view of FDR where $\theta_i$ are regarded as fixed and the expectation are taken with respect to the randomness induced by $(P_i)_{i\in\cH_0}$.  In this section we re-examine the multiple testing problem from a Bayesian perspective and review the SC procedure \citep{sun2007oracle} and its variations. 
\subsection{The SC procedure}
Suppose the underlying truths $\theta_i$ are random and the summary statistics $X_i$ are  modeled using the following hierarchical model
\begin{equation}\label{eq:bayes_model}
\theta_i\overset{\rm iid}{\sim}{\rm Bernoulli}(\pi),\quad X_i\overset{\rm ind}{\sim} (1-\theta_i)f_{0}+\theta_if_{1},\quad\forall i\in\cM,
\end{equation}
where the null distribution $f_{0}$ is assumed to be known. In the Bayesian framework, the expectation in the definition of FDR and mFDR is taken with respect to both $X_i$ and $\theta_i$. \cite{sun2007oracle} showed that the optimal testing procedure for controlling mFDR at level $\alpha$ is a thresholding procedure on the local false discovery rate (Lfdr), defined as
\begin{equation}\label{eq:lfdr}
	{\rm Lfdr}(X_i):=\PP(\theta_i=0|X_i)=\frac{(1-\pi)f_{0}(X_i)}{(1-\pi)f_{0}(X_i)+\pi f_{1}(X_i)}.
\end{equation}
For procedures that reject $H_{0,i}$ if and only if ${\rm Lfdr}_i\leq t$,
the SC procedure proposed by \cite{sun2007oracle} uses the following to estimate the FDP,
$$
\widehat{\rm FDP}^{\rm SC}(t)=\frac{\sum_{i=1}^{m}{\rm Lfdr}(X_i)\ind({\rm Lfdr}(X_i)\leq t) }{\sum_{i=1}^{m}\ind({\rm Lfdr}(X_i)\leq t)}.
$$
In practice,  parameters that comprise  the Lfdr values must be estimated from the data. The non-null proportion $\pi$ can be estimated using the methods mentioned in Section \ref{sec:adaBH}. The marginal distribution $(1-\pi)f_{0}(X_i)+\pi f_{1}(X_i)$ can be estimated using standard density estimation techniques \citep{ lindsey74a, lindsey74b, silverman2018density}. It can be shown that as long as the estimator $\widehat{\rm Lfdr}\overset{p}{\rightarrow} {\rm Lfdr}$ the SC procedure with ${\rm Lfdr}(X_i)$ replaced by its estimator $\widehat{\rm Lfdr}(X_i)$ in the definition of $\widehat{\rm FDP}^{\rm SC}(t)$ controls both FDR and mFDR and is optimal asymptotically. A key innovation of the SC procedure lies in its reliance on the Bayesian Lfdr rather than the frequentist p-value. This enables the development of an optimal multiple testing procedure that outperforms the optimal p-value based procedure, because it captures more information about the alternative distribution.

\subsection{SC Procedure with side information}
If in addition to the primary statistics $X_i$, a covariate $s_i$ that encodes side information is also available, the the knowledge from $s_i$ can be easily incorporated into the Lfdr statistic. 
Consider the following model for $X_i$
$$
\theta_i|s_i\overset{\rm iid}{\sim}{\rm Bernoulli}(\pi(s_i)),\quad X_i|s_i\overset{\rm ind}{\sim} (1-\theta_i)f_{0,s_i}+\theta_if_{1,s_i},\quad\forall i\in\cM.
$$
 Define the conditional Lfdr as 
\begin{equation}
	\label{eq:clfdr}
	\text{Clfdr}_i=\mathbb{P}(\theta_i=0|X_i,s_i)=\frac{(1-\pi(s_i))f_{0,s_i}(X_i)}{(1-\pi(s_i))f_{0,s_i}(X_i)+\pi(s_i) f_{1,s_i}(X_i)}.
\end{equation}
It was shown in \cite{fu2022heteroscedasticity} and \cite{tony2019covariate} that the optimal testing procedure for controlling the mFDR is a thresholding rule based on the Clfdr. If $\text{Clfdr}_i$ can be estimated consistently, then the SC procedure applied on $\widehat{\text{Clfdr}}_i$ can control both FDR and mFDR and is optimal asymptotically. This is indeed the approach taken in \cite{fu2022heteroscedasticity} and \cite{tony2019covariate}. 
 However, the $\text{Clfdr}_i$ is extremely difficult to estimate when the dimension of $s_i$ exceeds two. To address this problem \cite{leung2022zap} proposed to use a formula similar to \eqref{fdpbc} to estimate the FDP. More specifically, \cite{leung2022zap} used a beta-mixture model to construct an assessor function $a(\cdot,\cdot)$ to approximate the Clfdr. However, the FDR guarantee of the ZAP method proposed in \cite{leung2022zap} needs no assumption on the accuracy of the approximation.  Define $c_i(t)=\mathbb{P}(a(X_i,s_i)\leq t|\theta_i=0)$, then the FDP of tests  that reject $H_{0,i}$ when $a(X_i,s_i)\leq t$
 can be estimated by
 $$
 \widehat{\rm FDP}^{\rm zap}(t)=\frac{1+\sum_{i=1}^m\ind(c_{i}(a(X_i,s_i))\geq1-c_{i}(t))}{\sum_{i=1}^m\ind(a(X_i,s_i)\leq t)}.
 $$
 The key idea is to use $a(X_i,s_i)$ for ranking and the mirror sequence $c_{i}(a(X_i,s_i))$ for estimating the number of false rejections.  
 In practice, the beta-mixture model's parameters are derived from the data, leading to a dependency among the test
  statistics $a(X_i,s_i)$. Nonetheless,
 \cite{leung2022zap} showed that the beta-mixture model parameters converge towards certain fixed values.
 Consequently, the $a(X_i,s_i)$ values  can be regarded as asymptotically independent. A different tactic for confronting the challenges related to the accuracy of $\widehat{\text{Clfdr}}$ was investigated in \cite{gang2023unified}, where a formula similar to \eqref{eq3} is used to estimate the FDP.

\section{FDR Control under Dependence}\label{sec:depend}
Up to this point, our discussion has proceeded under the implicit assumption that the summary statistics $X_i$ are independent given the true state of affairs encapsulated by $\theta_i$. Nonetheless, real-world observations frequently deviate from such an assumption of conditional independence. While the BH procedure retains its capacity to control the false discovery rate (FDR) at the predetermined level under specific forms of dependence \citep{benjamini2001control,Finner2009OnTF,Ramdas2017AUT}, detailed below,  confronting general dependence requires innovative approaches to ensure the FDR is maintained. This section aims to present a overview of some of the latest advances related to dependence in the context of FDR control.


\subsection{Positive Regression Dependence Set (PRDS)}\label{sec:PRDS}
In this subsection, we focus on a specific type of dependence structure called a positive regression dependence set (PRDS), defined below. 
\begin{definition}\label{def:PRDS}
  A set $A\subseteq \mathbb{R}^m$ is said to be increasing if $\mathbf x\in A$ implies $\mathbf y\in A$ for all $\mathbf y\geq \mathbf x$. We say $\Xb=(X_1,\dots,X_m)$ has a positive regression dependence on the subset $\cI_0\subseteq\cM$ (PRDS) if for any $i\in \cI_0$ and increasing set $A\subseteq \mathbb{R}^m$, the function $x\mapsto \PP(\Xb\in A\mid X_i\leq x)$ is increasing.
\end{definition}
The PRDS property is a demanding form of positive dependence which can be interpreted, loosely speaking, as saying all pairwise correlations are positive.
 We remark that the original definition proposed in \citet{benjamini2001control} requires that the function $\PP(\Xb\in A\mid X_i= x)$ be increasing, which is stronger than the condition in Definition \ref{def:PRDS}. 
 This version of PRDS was first used by \cite{Finner2009OnTF}.
   The BH procedure controls FDR under PRDS dependence as the next theorem illustrates
 \begin{theorem}\label{thmprds}
 	If p-values $(P_i)_{i\in\cM:\theta_i=0}$ are PRDS, the BH procedure controls FDR at $\alpha |\cH_0|/m$.
 \end{theorem}
  Below we give two examples of PRDS dependence. 
\begin{example}
 	Let $\Xb=(X_1,\dots,X_m)$ be a random vector  form multivariate normal distribution $\cN(\mu,\Sigma)$. $\Xb$ is PRDS on $\cI_0\subseteq\cM$ if and only if $\Sigma_{ij}>0$ for all $i\in\cH_0$ or $j\in\cH_0$.
 	\label{eg:PRDS}
 \end{example}
 Consider testing $H_{0,i}:\mu_i=0$ versus $H_{1,i}:\mu_i>0$ for $\Xb\sim\cN(\mu,\Sigma)$ with $\Sigma_{i,j}\geq0$ for all $i,j\in\cM$. Following Example \ref{eg:PRDS}, one-sided p-values $P_i=\Phi(X_i)$ are PRDS on $\cH_0$ as $\Phi(x)$ is monotone increasing, thus allowing for valid application of the BH procedure.

 The next example is related to conformal inference \citep{shafer2008tutorial,bates2023testing,marandon2022machine,liang2024integrative}. A key concept in the conformal inference literature is the exchangeable distribution, which we define below.
 \begin{definition}
 	Consider random variables $Z_1,\ldots,Z_N$. Suppose that for any collection of $N$ values, the $N!$ different orderings are equally likely. Then we say that $Z_1,\ldots,Z_N$ are exchangeable.
 \end{definition}
 \begin{example}
 	Let $X_1,\ldots, X_n$ be samples from a null distribution $f_0$. Call these the training data. Further assume we have testing data $X_{n+1},\ldots, X_{n+m}$ such that $\{X_i: i=1,\ldots,n+m, \ X_i\sim f_0\}$ are exchangeable. Then it was shown in \cite{bates2023testing} that the conformal p-values $\{P_i\}_{i=1,\ldots,m, X_i\sim f_0}$ are PRDS, where
 	$$
 	P_i=\dfrac{1+|\{j=1,\ldots,n: X_j\leq X_i\}|}{n+1}.
 	$$
 \end{example}


Example 2 carries considerable importance because it facilitates the control of the FDR within the scope of novelty detection problems (a summary of research can be found in \cite{Pimenteletal2014-Novelty}), particularly in instances where the null distribution is not clearly defined or remains undetermined.


\subsection{Other Dependence Structures}
The PRDS condition is quite strong and often difficult to verify in practice. Nevertheless, FDR control under other forms of dependence has been widely studied in the literature. For instance, \cite{storey2004strong}, \cite{wu2008false}, and \cite{clarke2009robustness} have shown that, in the asymptotic sense, the BH procedure remains valid under weak dependence, Markovian dependence, and linear process models, respectively. Furthermore, \cite{cai2022laws} demonstrated that their weighted p-value procedure can control FDR asymptotically under certain forms of weak dependence.

Intuitively, if the dependence structure among tests is known, it can be exploited to aggregate weak signals from individual tests, thereby increasing the signal-to-noise ratio. This concept is explored in works such as \cite{benjamini2007false}, \cite{sun2009large}, and \cite{sun2011multiple}, where the incorporation of functional, spatial, and temporal correlations into inference methods has been shown to improve both the power and interpretability of existing techniques.

However, the studies by \cite{owen2005variance} and \cite{Finner2009OnTF} revealed that high correlation can lead to increased variability in testing outcomes. To mitigate this issue, \cite{leek2008general} and \cite{friguet2009factor} investigated multiple testing under factor models and demonstrated that subtracting common factors can substantially weaken the dependence structure. Additionally, \cite{efron2007correlation} and \cite{fan2012estimating} discussed methodologies for incorporating the dependence structure to obtain more accurate FDR estimates for a given p-value threshold. Lastly, \cite{du2023false} introduced an innovative approach that bypasses the p-value calculation step entirely, using mirror statistics for FDR control.

\subsection{General Dependence} \label{sec:BY}
The pioneering method for controlling the FDR under general dependence was introduced by \cite{benjamini2001control}. The authors demonstrated that the BH procedure can maintain FDR control  at level $\alpha S_m$ where $S_m=\sum_{i=1}^m \frac{1}{i} \approx \log m$. Based on these findings, a logical step is to recalibrate the initial BH target FDR level to   $\alpha/S_m$. The factor $S_m$ is known as the BY correction in the literature. 


Another methodological innovation,  termed the dependence-adjusted BH procedure (dBH), was put forward by \cite{fithian2022conditional} to control the FDR under general dependence. 
The key technical idea is to decompose the FDR according to the additive contribution of each hypothesis and use conditional inference to adaptively calibrate a separate rejection rule for each hypothesis to directly control its FDR contribution. More precisely, we can write the FDR as
$$
{\rm FDR}=\mathbb{E}\left(\dfrac{|\cR\cap\cH_0|}{|\cR|\vee 1}\right)=\sum_{i\in \cH_0}\mathbb{E}\left(\dfrac{\ind(p_i\leq\tau_i)}{|\cR|\vee 1}\right).
$$
The FDR can be controlled if we can ensure each individual summand in the above equation is bounded by $\alpha/m$. This is usually achieved by conditioning each term on a set of suitabe statistics to block the influence of most or all of the nuisance parameters on the conditional analysis. This conditional FDR calibration can be much more powerful than the BY correction, but this advantage comes at a practical cost. In general, conditional FDR calibration is more of a versatile theoretical blueprint than a practical algorithm. However, conditional FDR calibration turns out to be practical and intuitive to implement in integrative conformal p-values \citep{liang2024integrative}.

The final approach in this review of FDR control under arbitrary dependence acts through a concept called the e-value. 
\begin{definition}\label{eval}
An e-value is a non-negative random variable satisfying $\EE[E]\leq1$ under the null hypothesis.
\end{definition}
 If $E$ is an e-value, then, using the Markov inequality, it is straightforward to show that $1/E$ follows a super-uniform distribution, and hence is a special case of the p-value. Properties of the e-value and its relation to the p-value were  studied extensively in \cite{vovk2021values}.
  Given e-values $E_1,\ldots, E_m$ we can simply apply the BH procedure on $1/E_1,\ldots, 1/E_m$. Such a procedure is called e-BH. A significant breakthrough associated with the e-BH procedure, shown in \cite{wang2022false}, is its capability to control the FDR under conditions of arbitrary dependence among the e-values. 
  
A natural question is why one should care about e-values given the widespread use of p-values in the sciences. One important reason is that e-values can be constructed through martingales and betting scores, which arise naturally in sequential analysis \citep{wang2022false, shafer2021testing}. Moreover, e-values are more amenable to evidence aggregation \citep{vovk2021values}. For instance, when dealing with stages of data collection, if in the first stage we obtain an e-value $E_1$ and in the second stage another e-value $E_2$, the two e-values can be combined by simply taking their product, provided that  $\mathbb{E}(E_2|E_1) \leq 1$, or their arithmetic mean, providing a straightforward method for cumulative evidence integration. This is particularly advantageous in adaptive designs and online learning scenarios, where data arrive in streams and decisions need to be updated continually.

Additionally, by Ville's inequality \cite{ville1939etude}, e-values maintain their validity under optional stopping, a property that p-values do not generally possess. This means that researchers can conduct interim analyses and make decisions based on accumulating data without inflating the type I error rate, which is a significant benefit for real-time data monitoring and sequential testing frameworks. A more comprehensive discussion on the merits of e-values can be found in \cite{wang2022false}.

 If we are only interested in the multiple testing problem, then  $E_1,\ldots, E_m$ are only required to be generalized e-values for FDR control \citep{wang2022false,ren2022derandomized},
\begin{definition}
	\label{def:generalized_evalue}
	We say $(E_i)_{i\in\cM}$ is a set of generalized e-values if $\sum_{i\in\cH_0}\mathbb{E}[E_i]\leq m$. 
\end{definition}
One interesting fact about generalized e-values is that the decision sequence of a FDR controlling method can be converted into generalized e-values, which was  demonstrated in \cite{ren2022derandomized,banerjee2023harnessing} and \cite{li2023values}. More specifically, if a FDR controlling procedure is used to test $H_{0,1},\ldots, H_{0,m}$ at target FDR level $\alpha$ and the decision sequence is $(\delta_1,\ldots,\delta_m)$. Then it is straightforward to check that $\{e_i\}_{i=1}^m$ with $e_i=m\delta_i/(\alpha \sum_{i=1}^{m}\delta_i)$ is a generalized e-value. This fact can be leveraged for de-randomization \citep{ren2022derandomized,bashari2023derandomized} and meta-analysis \citep{banerjee2023harnessing}. 

In terms of power, neither e-BH nor BY correction strictly dominates the other. Both methods are quite conservative in practice. It is essential to recognize that e-values typically utilize less information compared to p-values. While p-values necessitate a complete specification of the distributions of the test statistics, e-values only rely on the information about a known mean. A comparison between e-BH and BY correction can be found in \cite{wang2022false}. Recently, \cite{xu2023more} showed that the power of e-BH can be improved via randomization. \cite{lee2024boosting} pointed out that it is possible to improve the power of e-BH through conditioning.
The field of e-values is quite new and rapidly evolving, and readers are encouraged to consult the latest literature for detailed discussions. The recent book \citep{ramdas2024hypothesis} provides an excellent reference for e-values.


\section{Discussions and Other Topics}
This article offers a survey of certain recent developments in the realm of FDR control. Nevertheless, we would like to highlight that this overview is not exhaustive and does not represent the full breadth of the latest progress in this field. 
Due to space constraints, we must omit the influence of FDR on problems other than multiple testing. For example, FDR has proven to be a highly useful concept in sparse vector estimation \citep{Abramovich2005AdaptingTU}, covariance matrix regularization  \citep{bailey2019multiple}, and 
graphical model estimation \citep{Liu2013GaussianGM}.
FDR has also found wide acceptance in variable selection problems \citep{Geer2013OnAO,barber2015controlling,barber2019knockoff,dai2023false,xing2023controlling}.
 Additionally, a series of recent papers pioneered the application of multiple testing schemes in machine learning problems such as multi-label classification \citep{angelopoulos2021learn,marandon2022machine}.

In this chapter we consider the problem of offline testing, where all summary observations are received at once, and all decisions must be made simultaneously. However, it is often at odds with modern data-driven decision-making processes requiring sequential decision-making. This has led to the development of online multiple testing procedures, which have been extensively studied in recent years \citep{foster2008alpha,aharoni2014generalized,ramdas2017online,ramdas2018saffron,gang2023structure}. Readers interested in online FDR control are encouraged to consult the review by \cite{robertson2022online}.

	
\bibliographystyle{apalike}
\bibliography{reference.bib}

\end{document}